\documentclass[%
 reprint,
 superscriptaddress,
 amsmath,amssymb,
 aps,
 prl,
 floatfix,
]{revtex4-1}

\usepackage{graphicx}
\usepackage{dcolumn}
\usepackage{bm}
\usepackage{braket}

\begin{document}

\title{Anharmonicity-driven Rashba co-helical excitons \\ break quantum efficiency limitation}

\author{Chang Woo Myung}
\email{cwmyung@unist.ac.kr}
\affiliation{Center for Superfunctional Materials, Department of Chemistry, Ulsan National Institute of Science and Technology (UNIST), 50 UNIST-gil, Ulsan 44919, Korea.}
\author{Kwang S. Kim}
\email{kimks@unist.ac.kr}
\affiliation{Center for Superfunctional Materials, Department of Chemistry, Ulsan National Institute of Science and Technology (UNIST), 50 UNIST-gil, Ulsan 44919, Korea.}
\affiliation{Center for Superfunctional Materials, Department of Physics, Ulsan National Institute of Science and Technology (UNIST), 50 UNIST-gil, Ulsan 44919, Korea.}


\begin{abstract}
Closed-shell light-emitting diodes (LEDs) suffer from the internal quantum efficiency (IQE) limitation imposed by optically inactive triplet excitons. Here we show an undiscovered emission mechanism of lead-halide-perovskites (LHPs) APbX$_3$ (A=Cs/CN$_2$H$_5$; X=Cl/Br/I) that circumvents the efficiency limit of closed-shell LEDs. Though efficient emission is prohibited by optically inactive $J=0$ in inversion symmetric LHPs, the anharmonicity arising from stereochemistry of Pb and resonant orbital-bonding network along the imaginary A$^+\cdots$X$^-$ (T$_{1u}$) transverse optical (TO) modes, breaks the inversion symmetry and introduces disorder and Rashba-Dresselhaus spin-orbit coupling (RD-SOC). This leads to bright co-helical and dark anti-helical excitons. Many-body theory and first-principles calculations affirm that the optically active co-helical exciton is the lowest excited state in organic/inorganic LHPs. Thus, RD-SOC can drive to achieve the ideal 50 $\%$ IQE by utilizing anharmonicity, much over the 25 $\%$ IQE limitation for closed-shell LEDs.
\end{abstract}

\maketitle

\begin{figure}
\centering
\includegraphics[width=8 cm]{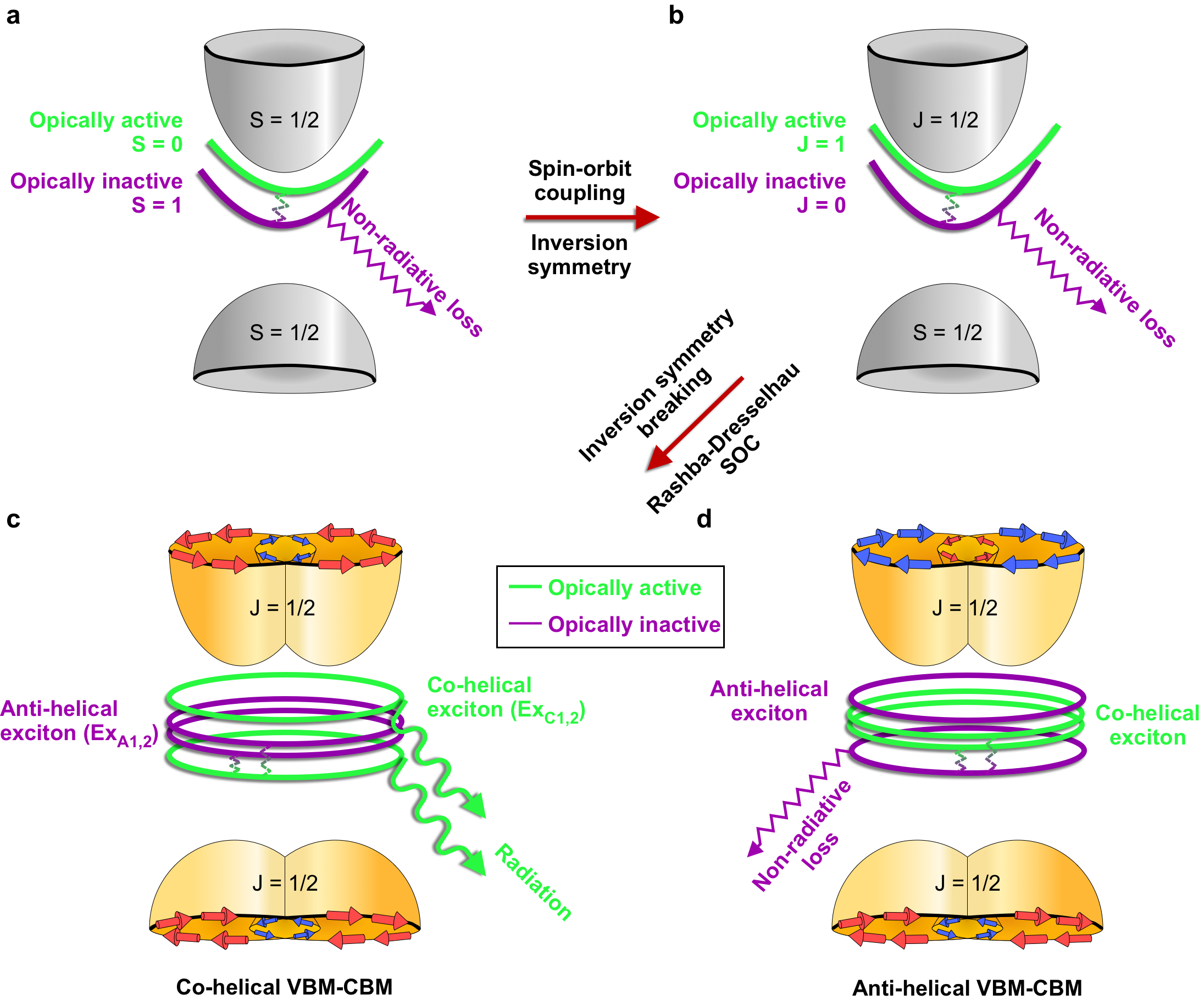}
\caption{(a) Without SOC, optically inactive triplet excitons ($S = 1$) are the lowest excited states. Generally, bright $S=0$ relaxes to dark $S=1$ via intersystem crossing (zig-zag line). (b) In lead halide perovskites, the strong SOC splits the triply degenerate p-type conduction bands into $J = 3/2$ (not shown) and $J = 1/2$. Still, the optically inactive singlet $J = 0$ exciton is the lowest excited state. When the inversion symmetry is broken, the RD-SOC introduces co-helical (Ex$_C$) and anti-helical (Ex$_A$) excitons. (c) When the helicities of CBM $\chi_{cbm}$ and VBM $\chi_{vbm}$ match ($\chi_{cbm} \cdot \chi_{vbm} = +1$), the optically active Ex$_{C1}$ becomes the lowest excited state, enabling an efficient PL. The energy ordering is Ex$_{C1}$ $<$ Ex$_{A1}$ $<$ Ex$_{A2}$ $<$ Ex$_{C2}$ (Supplementary Information). (d) When the $\chi_{cbm}$ and $\chi_{vbm}$ are opposite ($\chi_{cbm} \cdot \chi_{vbm} = -1$), the optically inactive Ex$_{A1}$ is the lowest excited state and becomes a channel for non-radiative energy loss with the energy ordering Ex$_{A1}$ $<$ Ex$_{C1}$ $<$ Ex$_{C2}$ $<$ Ex$_{A2}$.}
\label{fig:1}
\end{figure}

In recent years, a dossier of studies have reported high photoluminescence (PL) efficiencies using lead halide perovskite (LHP) light emitting diodes (PeLEDs) APbX$_3$(A=Cs$^+$, CN$_2$H$_5^+$ (Formamidinium or FA); X=Cl, Br and I)\cite{becker18, Cho1222, xu19, zhang17, ai18}, now achieving maximum external quantum efficiency over 21 \% \cite{xu19}. Closed-shell singlet LED materials suffer from the intrinsic efficiency loss (75 \%) because of optically inactive triplet excitons with an undesirable energy ordering \cite{becker18}. To overcome the limitation, various mechanisms have been pursued such as multiple exciton generations\cite{smith10}, defect-assisted PL\cite{piao13}  and doublet radical LED\cite{ai18}. Often, the lattice or valley degrees of freedom\cite{becker18, baldini18}, intermittency\cite{frantsuzov08}  and topological phase couples to the interesting excitonic properties\cite{cao18}. The Rashba-Dresselhaus (RD) spin-orbit coupling (SOC) arises in the solids containing heavy elements in which the inversion symmetry is violated, including LHPs\cite{etienne16,myung18,zhai17}. Recently, the crystal-glass duality of LHPs has been realized where the vibrational properties of LHPs are not well defined as either phonon (crystal) or amorphous medium (glass)\cite{miyata17,zhu19,pisoni14}. The lack of crystallinity (or disorder) of LHPs originates from the incommensurate octahedral tilting\cite{motta15}, fluctuations of A-site cation\cite{yaffe17}, anharmonic interactions\cite{miyata17,monserrat13,marronnier18,li15}, and strong phonon scatterings\cite{gold-parker18}. 

Here, we elucidate the mechanism for an efficient PL of organic/inorganic LHP by the combined effect of anharmonicity and RD-SOC using first-principles calculations and many-body theory at the PBE+D3 level, which is consistent with the PBE0+D3 level in this work. Because of the inversion breaking anharmonicity, the excitons are formed from the RD-split conduction band minimum (CBM) and valence band maximum (VBM). An intriguing result is that the RD spin helicities of CBM and VBM (or spin-pair helicity) match along all the imaginary TO displacements and the co-helicity is invariant, thereby forming the bright co-helical exciton (E$_{xC}$) as the lowest excited state harnessing excitonic emissions efficiently\cite{kasha50}. 

In general, the excited state of singlet LEDs emits photons from an optically active singlet ($S = 0$), while the supplied energy is dissipated through optically inactive triplets ($S = 1$) non-radiatively (Fig. 1a). In addition, the energy ordering of exciton spin multiplets is usually unfavorable because of the exchange splitting (Fig. 1a). Under the strong SOC by heavy elements, optically active $J = 1$ and optically inactive $J = 0$ are formed (Fig. 1b). This is the case for inversion symmetric inorganic LHPs (Fig. 1b). However, the energy ordering is still unfavorable for an efficient PL due to exchange splitting. If the inversion symmetry is violated under strong SOC, the RD-SOC follows by splitting the $J = 1/2$ band into two helical spin conduction(valence) bands $\chi_{c(v)}$. For instance, the Rashba type interactions lead to $\chi_{c(v)} = +1$ (counter-clockwise) and $-1$ (clockwise) (Fig. 1c, d). Then, the excited states from the helical RD conduction and valence bands result in either bright co-helical excitons (Ex$_{C}$), $\chi_{cbm} \cdot \chi_{vbm}=+1$, or dark anti-helical excitons (Ex$_{A}$), $\chi_{cbm} \cdot \chi_{vbm}=-1$. A compelling observation is that an advantageous energy ordering of excitons can be obtained when the spin-pair helicity of CBM and VBM match, forming Ex$_{C1}$ as the lowest excited state (Fig. 1c and Supplemental Material). Otherwise, Ex$_{A}$ becomes the source of non-radiative losses as a result of helicity-mismatch (Fig. 1d and Supplemental Material). 

\begin{figure*}
\centering
\includegraphics[width=16 cm]{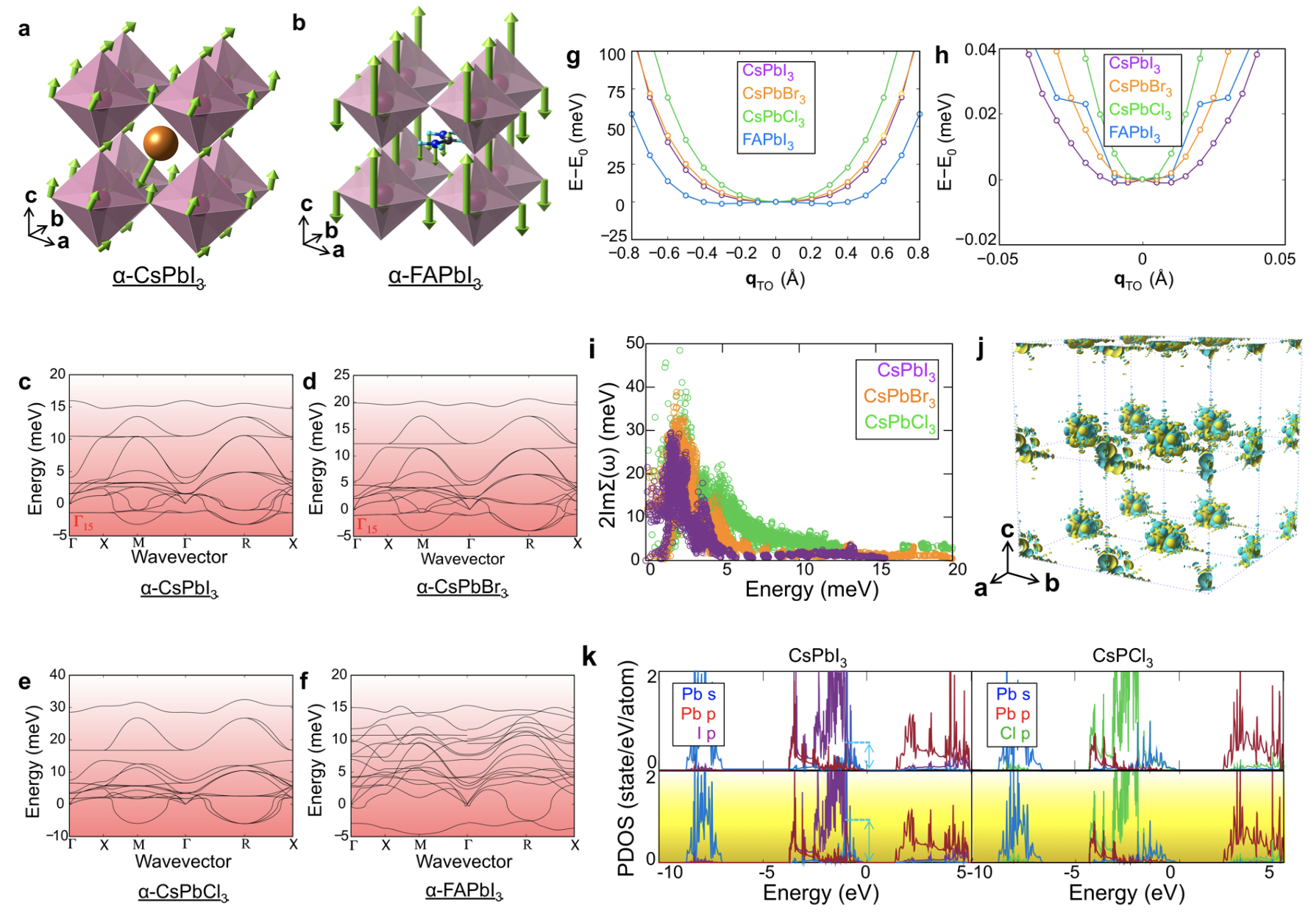}
\caption{(a) T$_{1u}$ TO$_i$ $\Gamma_{15}$ modes in $\alpha$-CsPbI$_3$ propagating along $a$. (b) Imaginary TO mode propagating along $a$ in $\alpha$-FAPbI$_3$ (eigenmode: green arrows). (c)-(f) Phonon dispersions (black line) of $\alpha$-CsPbX$_3$ (X=I/Br/Cl) and $\alpha$-FAPbI$_3$ with [100]-oriented FA. (g) PES along TO$_i$ mode for $\alpha$-CsPbX$_3$ at the PBE-D3 level (consistent with the PBE0-D3 results). E$_0$ is E($q = 0$). (h) Enlarged image of the minima region of (g). (i) Phonon lifetime $\tau^{-1}(\omega)= 2Im\Sigma (\omega)$ at 300 K for $\alpha$-CsPbX$_3$. (j) Visualization of ($10 \times 10 \times 10$) supercell of $\alpha$-CsPbI$_3$ with charge density perturbation $\Delta \rho$ upon TO$_i$ mode with 10$^{-5}$ e/\AA$^3$ iso-surface. (k) Partial density of states of $\alpha$-CsPbI$_3$ and $\alpha$-CsPbCl$_3$ at equilibrium (up) and upon TO$_i$ eigenmode displacement (0.1 \AA) (down, yellow). The Pb-$6s$ and -$6p$ hybridization (cyan arrows) increases only in $\alpha$-CsPbI$_3$ with TO$_i$ displacement.}
\label{fig:2}
\end{figure*}

The inversion symmetry breaking anharmonicity in LHP originates from the low energy $T_{1u}$ TO motion (TO$_i$) of A-site cation and halide X (Fig. 2a, b) identified by density functional perturbation theory calculations (DFPT) giving E$_{\Gamma}(\alpha$-CsPbI$_3$) = 1.414$i$ meV, E$_{\Gamma}(\alpha$-CsPbBr$_3$) = 1.09$i$ meV, E$_{\Gamma}(\alpha$-FAPbI$_3$) = 2.95$i$ meV (Fig. 2c-f), consistent with the previous studies\cite{miyata17,monserrat13,marronnier17}. 

\begin{figure}
\centering
\includegraphics[width=8 cm]{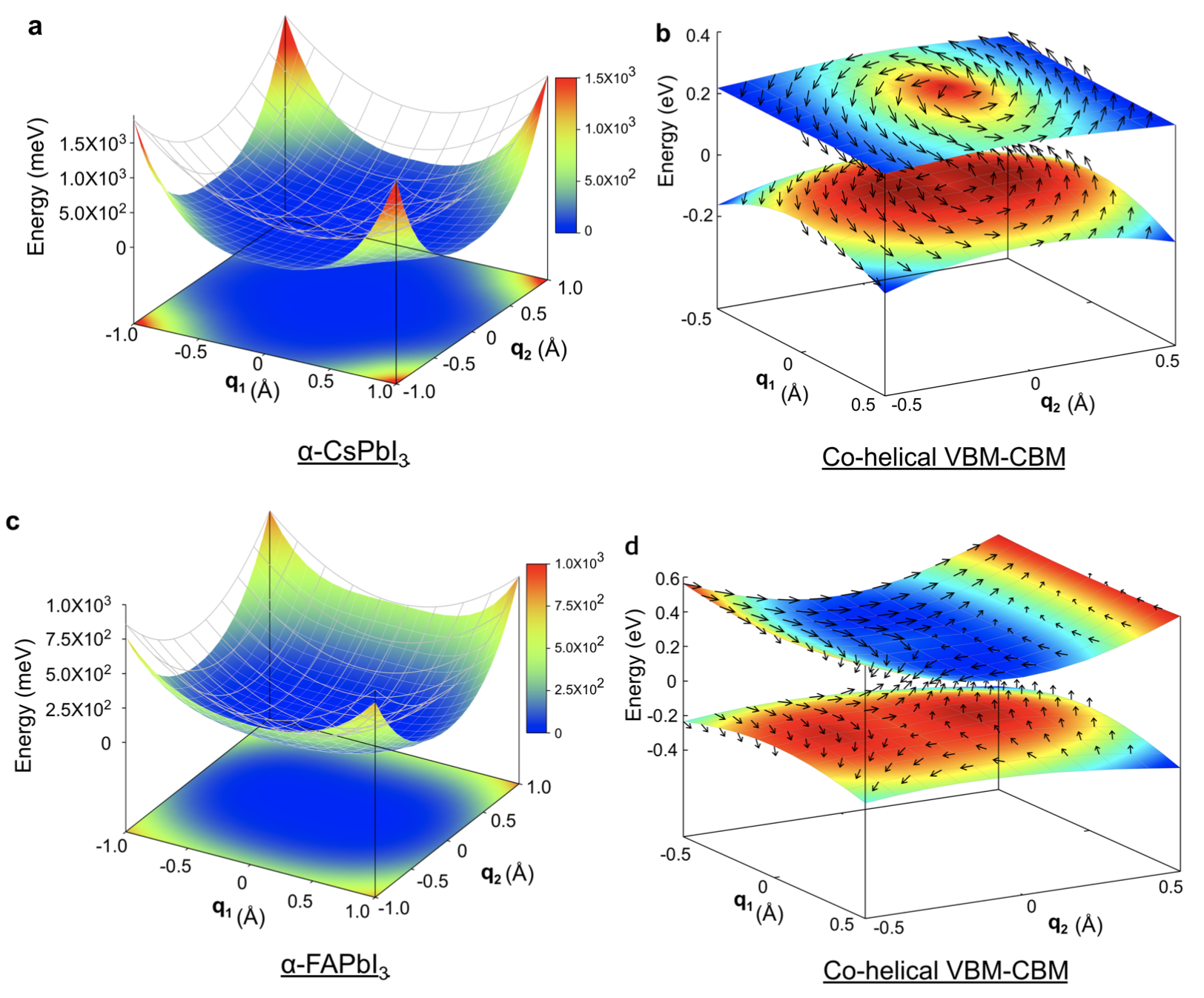}
\caption{The 2D PES map (in meV) of frozen phonon displacements $q_1$ and $q_2$ (-1.0 \AA $ \leq q_1,q_2 \leq $ 1.0 \AA) in (a) $\alpha$-CsPbI$_3$ and (c) $\alpha$-FAPbI$_3$. Harmonic PESs (gray-grid surface) manifest strong higher order anharmonicity of TO$_i$ modes in organic/inorganic LHPs. Non-collinear magnetization vectors ($m_x, m_y, m_z$) and energies of CBM and VBM along $q_1$ and $q_2$ (-0.5 \AA $ \leq q_1,q_2 \leq $ 0.5 \AA) of the TO$_i$ modes in (b) $\alpha$-CsPbI$_3$ and (d) $\alpha$-FAPbI$_3$. The parallel magnetization vectors at each TO$_i$ displacement ($q_1$, $q_2$) indicate the co-helicity between the bands.}
\label{fig:3}
\end{figure}

\begin{figure}
\centering
\includegraphics[width=8 cm]{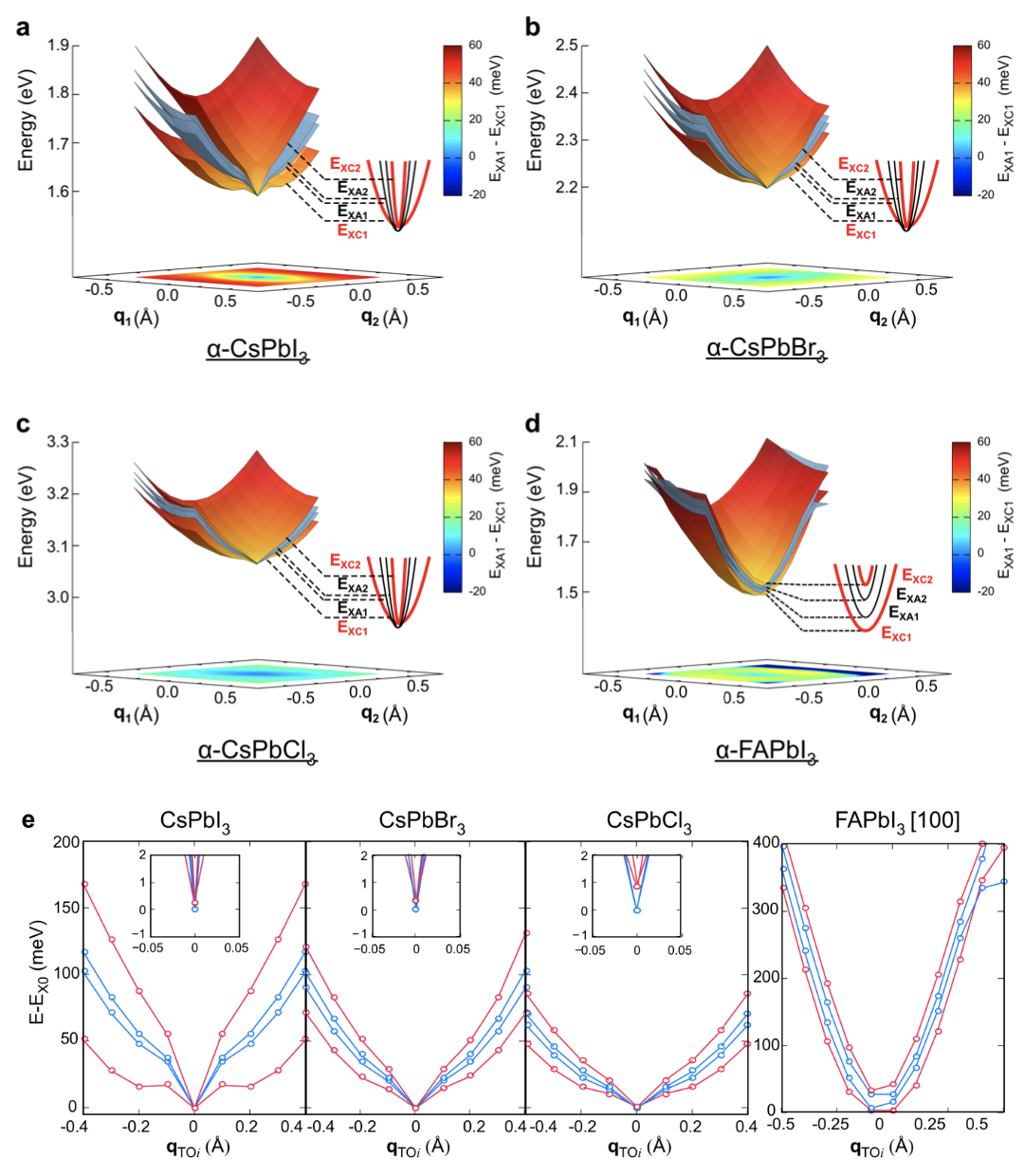}
\caption{The BSE co-helical Ex$_{C1}$, Ex$_{C2}$ (red surface) and anti-helical Ex$_{A1}$, Ex$_{A2}$ (blue surface) excitons energy levels and the mapping of their energy splitting $\Delta E=E_{Ex_{A1}}-E_{Ex_{C1}}$ in (a) $\alpha$-CsPbI$_3$, (b) $\alpha$-CsPbBr$_3$, (c) $\alpha$-CsPbCl$_3$ and (d) $\alpha$-FAPbI$_3$ ([100] FA direction) along the frozen phonon displacements q$_1$ and q$_2$ of TO$_i$ (-0.5 \AA $\leq q_1,q_2 \leq $ 0.5 \AA). (e) Energy crossing between Ex$_{C1,2}$ (red) and Ex$_{A1,2}$ (blue) excitons around $q_1=q_2 \sim 0$ in inorganic LHPs. Except for around $q_1=q_2 \sim 0$ where the dark $J = 0$ is the lowest excited state, the TO$_i$ displacements allow the bright Ex$_{C1}$ to be the lowest level. In $\alpha$-FAPbI$_3$, Ex$_{C1}$ is the lowest level at $q = 0$, because FA perturbs the inversion symmetry dynamically and leads to RD-SOC.}
\label{fig:4}
\end{figure}

The interaction between A$^+$ and X$^-$ is mostly ionic but contains a sizable contribution of dispersion interactions in organic LHP from the CCSD(T) calculations (Supplemental Material). Though the anharmonicity of both TO$_i$ modes (inversion symmetry broken) and rotational acoustic modes (inversion symmetric)\cite{gold-parker18} are strong, only anharmonic TO$_i$ modes can create Ex$_{C}$ by breaking the inversion symmetry ({\textit Pm3m} $\to$ {\textit Pm}) coupled to RD-SOC. The TO$_i$ motion is identical for inorganic LHPs except for the amplitude of individual atom (Supplemental Material). The nature of anharmonicity of LHP is investigated by calculating expansion of the Hamiltonian, $\mathcal{H}=\mathcal{H}_2+\mathcal{H}_3+\mathcal{H}_4 = \frac{1}{2!}\sum_{ij\alpha \beta}\Phi^{\alpha \beta}_{ij} +  \frac{1}{3!}\sum_{ijk\alpha \beta \gamma}\Phi^{\alpha \beta \gamma}_{ijk} +  \frac{1}{4!}\sum_{ijkl\alpha \beta \gamma \epsilon}\Phi^{\alpha \beta \gamma \epsilon}_{ijkl}$, in terms of atomic displacement $u_{\alpha i}$ of $i$-th atom along the $\alpha$ direction, where we calculate $\Phi^{\alpha \beta \gamma}_{ijk}$ by using the finite displacement method and $\Phi^{\alpha \beta \gamma \epsilon}_{ijkl}$ by fitting the potential energy surface (PES).

The fourth order anharmonicity $\Phi^{\alpha \beta \gamma \epsilon}_{ijkl}$, responsible for almost flat PES of TO$_i$, is pronounced in the order I $>$ Br $>$ Cl (Fig. 2c-e, 2g, Supplemental Material). The PES within $q \sim 0.5$ \AA{} is almost flat ($<$ 25 meV) so that the TO$_i$ fluctuates without any loss of energy at room temperature, which is the source of disordered fluctuations. The lack of fourth order anharmonicity or the resulting disorder of $\alpha$-CsPbCl$_3$ (Fig. 2e, Supplemental Material) is associated with the poor PL efficiency of Cl-based PeLED\cite{kumawat19}  compared to I- or Br-based PeLEDs\cite{becker18,xu19}, which will be elaborated below. When FA is present, the TO$_i$ modes persist while one of the doubly degenerate TO modes is accentuated compared to $\alpha$-CsPbI$_3$ (Fig. 2f). A complete PES of $\alpha$-FAPbI$_3$ is highly anharmonic because of a large degree of freedom in FA’s orientation. The large double well minima (Fig. 2g) of $\alpha$-FAPbI$_3$ is caused by a slight tilting of FA. TO$_i$ motions in $\alpha$-FAPbI$_3$ show a sizable anharmonicity (Fig. 2g) regardless of FA dipole directions (Supplemental Material). The results reflect the universal strong fourth order anharmonicity of organic/inorganic LHPs. We note that there exist very small local minima where the PES seems almost flat and these local minima are responsible for the TO$_i$ modes appearing in DFPT calculations (Fig. 2c-f). This can be recognized with a much finer grid of TO$_i$ displacements. We find that a very small double well PES exists in $\alpha$-CsPbI$_3$ (at 0.01 \AA{} displacement) and $\alpha$-CsPbBr$_3$ (at 0.005 \AA{} displacement) (Fig. 2h), showing the limitation of finite displacement method by which TO$_i$ modes are not realized\cite{zhu19} with large displacements $\sim$ 0.1 \AA. We note that the anharmonic PESs within $q \sim$ 0.5 \AA{} displacements are almost flat ($\Delta E < 25 meV$) regardless of the different level of DFT exchange-correlation (XC) (Supplemental Material). The low optical modes also dominantly contribute to the third order anharmonicity $\Phi^{\alpha \beta \gamma}_{ijk}$ in $\alpha$-CsPbX$_3$, revealing that three phonon scattering  $\tau^{-1}=2 \text{Im} \Sigma(\omega)$ increases in the order I $<$ Br $<$ Cl (Fig. 2i and Supplemental Material). 

Furthermore, the origin of anharmonicity of TO$_i$ is understood as the coupled effect of the resonant bonding of weakly $sp$ hybridized orbitals (in which the sizes of interatomic force constants are significant over several unit-cells away\cite{zhu19, li15}; Fig. 2j and Supplemental Material) and the stereochemically active Pb $6s$ lone pair that hybridizes with Pb 6p via X $np$ upon TO$_i$ motions (Fig. 2k). Contrary to $\alpha$-CsPbI$_3$, the hybridizations of Pb $6s$ and $6p$ by TO$_i$ displacements are not allowed because of high energy levels of Pb $6p$ in $\alpha$-CsPbCl$_3$ (Fig. 2k). 

The effective RD-SOC along the TO$_i$ displacements u$_{1,2}$ is given by $\hat{H}_{SOC}= \lambda \sigma \cdot \{ p \times \nabla V(u_1, u_2) \} $, where $\lambda$, $\sigma$, $p$ and $\nabla V$ are SOC parameter, spin operator, momentum operator, and the electric field by inversion symmetry breaking, respectively\cite{manchon15}. Depending on the crystal symmetries, the SOC interactions take different forms: Dresseulhaus SOC,  $\hat{H}_D \propto \{ (p^2_y - p^2_z)p_x\sigma_x + c.p. \}$ in the zinc-blende crystals or Rashba SOC $\hat{H}_R \propto (z\times p)\cdot \sigma$ in the interfacial asymmetry along the z direction where c.p. refers to the circular permutations of indices\cite{manchon15}. Because anharmonic TO$_i$ fluctuations have low energy barrier, TO$_i$ motions u$_{1,2}$ (Fig. 3a, c and Supplemental Material) are the dominant source of RD-SOC (Fig. 3b, d and Supplemental Material)\cite{monserrat17}. Under a harmonic motion $u$, one of the split $J = 1/2$ bands contains the equal contributions of opposite displacements $+u$ and $-u$. The opposite displacements lead to the mixing and cancellation of spin eigenstate of RD interactions by $\bra{\psi^{-u}_k} \hat{J} \ket{\psi^{-u}_k} = \bra{\psi^{+u}_k} \hat{T}^{-1} \hat{I}^{-1} \hat{J} \hat{I} \hat{T} \ket{\psi^{+u}_k} = \bra{\psi^{+u}_k} -\hat{J} \ket{\psi^{+u}_k} $, where $\hat{I}$ and $\hat{T}$ are inversion and time-reversal operator\cite{kim14}. Thus, the subsequent band states remain as two split effective $J = 1/2$ bands, which lead to an inefficient PL (Fig. 1b). However, we need to elucidate if the spin of the subsequent exciton state mix into $J = 1/2$ by TO$_i$ displacements in average. To make an efficient LED, $\chi_{cbm}$ and $\chi_{vbm}$ must be always co-helical $\chi_{cbm} \cdot \chi_{vbm} = +1$  (protected from the spin mixing) so that the optically bright Ex$_{C}$ is the lowest excited state (Fig. 1c). Although the prediction of spin-pair helicity of CBM and VBM is difficult, there exist some computational results that halide fluctuations contribute to the co-helicity, while Pb motions are related to the anti-helicity\cite{kim14,zheng15}. We elaborate that the lowest exciton spin state is RD co-helical and is invariant to the TO$_i$ vibrations for LHPs in the following.

The individual spin helicity of $\chi_{cbm}$ and $\chi_{vbm}$ dynamically changes its helicity under TO$_i$ vibrations\cite{kim14}, but we find that the non-collinear helical spins of CBM and VBM have the spin-pair helicity of  $\chi_{cbm} \cdot \chi_{vbm} = +1$ under the TO$_i$ displacements in $\alpha$-CsPbI$_3$ (Fig. 3b). The non-collinear spin states with TO$_i$ displacements in $\alpha$-CsPbBr$_3$ and $\alpha$-CsPbCl$_3$ demonstrate the consistent co-helicity between CBM and VBM (Supplemental Material) regardless of the halide type in inorganic LHPs. We also find that the non-collinear spins of CBM and VBM in $\alpha$-FAPbI$_3$ (Fig. 3d) are co-helical. The co-helicity of CBM and VBM in $\alpha$-FAPbI$_3$ is invariant to the different FA dipole directions [110] and [111] (Supplemental Material). The results manifest that ($\chi_{cbm} \cdot \chi_{vbm}$) is protected to be co-helical under the anharmonic TO$_i$ vibrations, irrespective of halide types in $\alpha$-CsPbX$_3$ (Fig. 3b and Supplemental Material) and A-site cation directions in $\alpha$-FAPbI$_3$ (Fig. 3d and Supplemental Material). Therefore, we expect that the bright Ex$_{C}$ is always the lowest exciton level. On the contrary, ($\chi_{cbm} \cdot \chi_{vbm}$) along the lowest T$_{1u}$ LO of dominant Cs$\cdots$Pb motion is found to be anti-helical (Supplemental Material).

The optically inactive singlet $J = 0$ is lower than the optically active triplet $J = 1$ only at the inversion symmetric \textit{Pm3m} crystallographic position in inorganic $\alpha$-CsPbX$_3$. Except for this position, the subsequent energy levels of Ex$_{C}$ and Ex$_{A}$ obtained by the Bethe-Salpeter equation (BSE) calculations demonstrate that Ex$_{C}$ is always the lowest excited energy level under the TO$_i$ fluctuations irrespective of the halide and A-site cation type (Fig. 4 and Supplemental Material). Thus, there exists the intersystem crossing $\ket{j=0} \leftrightarrow \ket{Ex_C}$ around $q = 0$ for all inorganic LHPs (Fig. 4a-c, e). The energy level splitting between Ex$_C$ and Ex$_A$ is in the order of Cl $<$ Br $<$ I (Fig. 4a-c, 4e), which follows the SOC strength of Cl $<$ Br $<$ I. The exchange splitting between $J = 1$ and $J = 0$ excitons at $q_{TO_i}=0$ is in the order of Cl $<$ Br $<$ I (Fig. 4a-c, 4e). 

In $\alpha$-FAPbI$_3$, however, the energy crossing between Ex$_C$ and Ex$_A$ occurs at large displacements of TO$_i$, while the Ex$_C$ is the lowest energy level at $q_{TO_i}=0$ due to the inversion breaking of FA molecule and the following incommensurate octahedral distortions (Fig. 4d, e). The favorable energy ordering Ex$_{C1}$ $<$ Ex$_{A1}$ is retained regardless of FA dipole direction of [110] and [111] (Supplemental Material), which explicitly manifests the advantage of organic A-site cation by its local inversion symmetry breaking in PeLED. 

Our study reveals the undiscovered mechanism of co-helical versus anti-helical excitons in PeLED by the RD-SOC coupled to the anharmonic glassy disorder. This achieves the ideal 50 \% IQE by circumventing the 25 \% IQE limitation of conventional closed-shell singlet LEDs, paving a way to highly efficient next-generation emitters. 

\begin{acknowledgments}
K.S.K. acknowledges the support from NRF (National Honor Scientist Program: 2010-0020414). K.S.K. acknowledges the support from KISTI (KSC-2019-CRE-0103, KSC-2019-CRE-0253, KSC-2020-CRE-0049). C.W.M. acknowledges the support from KISTI (KSC-2018-CRE-0071, KSC-2019-CRE-0139, KSC-2019-CRE-0248). C.W.M. thanks Lihua Wang for helpful discussions.
\end{acknowledgments}

\bibliography{apssamp}

\end{document}